\documentclass [12pt]{article}

\begin{document}

\begin{center}
{\Large On physical meaning of Weyl vector}
\end{center}

\begin{center}
\textbf{M.V.Gorbatenko}
\end{center}

\begin{center}
Russian Federal Nuclear Center -- VNIIEF, 

Sarov, N.Novgorod Region, Russia; E-mail: \underline{gorbatenko@vniief.ru}
\end{center}

\textbf{Abstract}

The present paper considers if the new proposed conformal geometrodynamics 
(CGD) can extend the Nature features compared with general theory of 
relativity (GTR). The answer for this question can be connected with unique 
phenomenon arising from Riemann space transition used in GTR, to Weyl space 
used in CGD. We have in mind the possibility to set up the perfect 
correspondence in certain spatial regions between the equations of different 
physical phenomena: (1) Phenomena associated with Weyl degrees of freedom in 
a plane space. (2) Phenomena described in terms of half-integer spin 
particles and observed quantities corresponding to the full set of 
bispinors. The said phenomenon is described in the present paper. Analyzed 
here are the new prospects in the problem of combination of quantum physics 
concepts and GTR, unification of physical interactions and understanding of 
many of the effects known from the experiments but not properly understood 
yet.

\section{Introduction }
\label{sec:introduction}

The general theory of relativity (GTR) is used in modern science and 
technology from Global Positioning System to cosmological models to describe 
space-time relations between the events. This is not by accident because of 
number of direct experimental correctness proofs based on GTR (see reviews 
in \cite{Sha}). It turned out that whenever there is a 
possibility to set up an experiment to check GTR, predictions of GTR met 
experimental results in the best way rather than alternative theories. 

However, the most convincing GTR check is confined to a weak gravitational 
field region, or as they usually say, post-Newtonian approximation (more 
accurate in some cases). In regard to high gravitational fields, modern 
science encounters considerable difficulties. The physics of the processes 
in material and space-time in the region of event horizons and singularity 
is still obscure. The attempts to develop energy-momentum tensor $T_{\alpha 
\beta}$ of the right side of GTR\footnote{ From now on we shall use 
notations from the known book in GTR \cite{Whe}.} equation 
led to no result

\begin{equation}
\label{eq1}
R_{\alpha \beta } - \frac{1}{2}g_{\alpha \beta } R = T_{\alpha \beta } ,
\end{equation}

\noindent
from considerations of geometry. The attempts combine the most successful 
theories of the 20$^{th}$ century -- GTR and quantum theory. 

Given the state of space-time science, the attempts to fall outside the 
limits of GTR are not insensible. One of the variants of such a fall was 
suggested and developed in the series pf papers 
\cite{GorbPush1984}-\cite{Gorb2005} et al. By 
conformal geometrodynamics (CGD) we shall call the specified theory. CGD 
equations are given by (\ref{eq1}), only $T_{\alpha \beta } $ tensor has been never 
used as energy-momentum tensor of the particular material. $T_{\alpha \beta 
} $ tensor is written as $A_\alpha $ vector and $\lambda $ scalar function 
in CGD as follows:

\begin{equation}
\label{eq2}
T_{\alpha \beta } = - 2A_\alpha A_\beta - g_{\alpha \beta } A^2 - 2g_{\alpha 
\beta } A^\nu _{;\nu } + A_{\alpha ;\beta } + A_{\beta ;\alpha } + g_{\alpha 
\beta } \lambda .
\end{equation}

The word ``conformal'' in reference to geometrodynamics is connected with 
the fact that equations (\ref{eq1}) with right side of in the form of tensor (\ref{eq2}) 
show invariance as regard to transformations

\begin{equation}
\label{eq3}
g_{\alpha\beta}\rightarrow g_{\alpha\beta}\cdot \exp\left(2\sigma\right),\ A_\alpha\rightarrow A_\alpha-\sigma_{;\alpha},\ \lambda\rightarrow\lambda\cdot\exp\left(-2\sigma\right).
\end{equation}

\noindent
Here $\sigma(x)$ - is 
coordinate arbitrary scalar function. Metrics transformations by law (\ref{eq3}) are 
called conformal. From the point of view of physicist, the transformations 
(\ref{eq3}) are of considerable value as they maintain cause-and-effect relations 
between the events. Mathematicians put this property of transformations (\ref{eq3}) 
in other words: transformations (\ref{eq3}) maintain the variety of light cone 
congruence. 

``Geometrodynamics'' term used in CGD abbreviation is explained by the 
possibility to set up the equation (\ref{eq1}) with right side in the form of tensor 
(\ref{eq2}) only in terms of geometric objects of Weyl space:

\begin{equation}
\label{eq4}
g_{\alpha\beta}=-{{1}\over{\lambda}}\cdot{\Re}_{\left(\alpha\beta\right)}.
\end{equation}

\noindent
Here ${\Re}_{\left(\alpha\beta\right)}$ - is 
symmetrical part of Ricci tensor for Weyl space. We are not going to specify 
Weyl space properties given in multiplicity of works beginning with 
\cite{Weyl}. Note that $A_\alpha$ vector has geometrical-only meaning as it is Weyl 
vector entering into connection structure

\begin{equation}
\label{eq5}
{\Gamma_\alpha^\mu}_\nu={{\nu}\choose{\alpha\nu}}+A_\alpha \delta^\mu_\nu+A_\nu \delta^\mu_\alpha-g_{\alpha\nu} g^{\mu\epsilon}A_\epsilon.
\end{equation}

\noindent
However, covariant derivatives of Weyl space object are found by the formulae 

\begin{equation}
\label{eq6}
\nabla Y^\beta=Y^\beta_{,\alpha}+\Gamma_\alpha^\beta{}_\epsilon Y^\epsilon+n\cdot A_\alpha Y^\beta .
\end{equation}

\noindent
Here $n$ - is Weyl 
object weight ($Y^\beta$ vector in that case).

The number of CGD equation nontrivial properties and solutions is described 
in [6]. Let us enumerate these properties:
\begin{enumerate}
\item
Cauchy problem for equation (\ref{eq1}) with right side in the form of tensor (\ref{eq2}) is 
set without Cauchy data.
\item
Space-like hypersurface crack solutions are admitted.
\item
There occurs preserving current vector along with a possibility to 
``process'' $T_{\alpha\beta}$ tensor 
using phenomenological thermodynamics. 
\item
Geometrodynamical medium proved to be considered as relativistic simple 
viscous liquid. Equations of state for this medium, as well as viscosity 
result from CGD equations. There occurs a new state function analogous to 
entropy.
\item
Gauge vector and lambda term can be interpreted in terms of the degree of 
freedom of $\raise.5ex\hbox{$\scriptstyle 1$}\kern-.1em/ 
\kern-.15em\lower.25ex\hbox{$\scriptstyle 2$} $ spin particles. 
\end{enumerate}

Several solutions of the equations of CGD have been obtained in recent 
years, the analysis of these equations has verified the enumerated 
properties (See \cite{Gorb2005}, \cite{Gorb2003}-\cite{GorbKoch}). It is proved, 
for example, that all known exact solutions of GTR equations (exterior and 
interior solutions of Schwarzschild, de Sitter, Friedmann etc.) can be 
arbitrary closely approximated in a certain space-time region by CGD 
solutions. Thus, it is reasonable to suppose that not only GTR but also CGD 
can be used to describe large-scale physical processes.

Either of the given above properties is interesting in its own way, and is 
likely to be analyzed. The most intriguing is the last property. For the 
first time geometrical objects proved to be connected with dynamics of 
operator of state of microscopic objects with half-integer spin. None of the 
attempts of physical interaction unification gave such a result. Thus, it is 
no wonder that it is the 5$^{th}$ property that is thoroughly analyzed. The 
algorithm of tensor system mapping onto bispinor degrees of freedom 
developed in a number of works (\cite{GorbPush2001}, \cite{GorbPush1999}, etc.) was used. The results of 
\cite{Gorb2009} research exceeded all expectations. Outline of 
the results is the aim of this work. 

The work includes auxiliary sections 2, 4, 5. The first one gives formulae 
of thermodynamic phenomenological analysis of geometrodynamic analysis of 
the medium described by energy-momentum tensor (\ref{eq2}). Section 4 contains 
well-known information of the theory of Dirac matrix. Section 5 gives the 
variant of tensor system mapping onto bispinor degrees of freedom. The new 
results are presented in sections 3, 6. Section 7 gives one of the possible 
exact solutions of CGD equations -- solution of Yukava potential type. The 
given solution is particular, but suffices to describe methods of 
application of the new results. The discussion of the results obtained is 
given at the end of the work.

\section{Thermodynamic analysis of CGD equations}
\label{sec:thermodynamic}

It follows from (\ref{eq28}) that vector $j_\alpha=Const\cdot\left(\lambda_{;\alpha}-2\lambda A_\alpha\right)$ 
generally satisfies the equation of continuity

\begin{equation}
\label{eq7}
j^\alpha_{;\alpha}=0.
\end{equation}

 \noindent
The timelike vector $j^\alpha$ 
can be represented in the space-time domain in the form of

\begin{equation}
\label{eq8}
j^\alpha\equiv\rho u^\alpha,
\end{equation}

\noindent
where $u^\alpha$ - is the unit 
timelike vector. In signature $\left(-+++\right)$

\begin{equation}
\label{eq9}
u^2=-1.
\end{equation}

The timelike vector $j^\alpha$ in the scheme satisfying the equation of continuity (\ref{eq7}), means that 
the scheme contains certain strongly conserved substance. The density 
$\rho$ of this substance 
is defined by the formula $\rho=\sqrt{-\left(j^\alpha j_\alpha\right)}$ from (\ref{eq8}), (\ref{eq9}).

Let us take a conserved substance to imply some charge which is taken for 
strongly conserved in elementary particle theory. The specific volume 
$V$ is defined as a 
reciprocal of $\rho$,

\begin{equation}
\label{eq10}
V={{1}\over{\rho}}.
\end{equation}

Two projection operators can be set up in the ordinary way with the help of $u^\alpha$:

\begin{equation}
\label{eq11}
-u^\alpha u^\beta, \, s^{\alpha\beta}\equiv g^{\alpha\beta}+u^\alpha u^\beta.
\end{equation}

The tensor $T_{\alpha\beta}$ in (\ref{eq2}) can be represented in the form of

\begin{equation}
\label{eq12}
T_{\alpha\beta}=U\cdot u_\alpha u_\beta +\left(u_\alpha q_\beta+u_\beta q_\alpha\right)+W_{\alpha\beta},
\end{equation}

\noindent
where the values $U,q_\alpha,W_{\alpha\beta}$ are defined by the formulae

\begin{equation}
\label{eq13}
U\equiv \left(u^\mu T_{\mu\nu} u^\nu\right),\, q_\alpha\equiv -s_\alpha {}^\mu T_{\mu\nu} u^\nu, \, W_{\alpha\beta}\equiv s_\alpha {}^\mu s_\beta {}^\nu T_{\mu\nu},
\end{equation}

Henceforth we shall follow (\ref{eq13}), that is to say: $U$ - is the energy density, 
$q_\alpha$ - is the energy flux vector, $W_{\alpha\beta}$ - is 
the strain tensor. $W_{\alpha\beta}$ tensor is often represented as the sum of two summands,

\begin{equation}
\label{eq14}
W_{\alpha\beta}=P\cdot s_{\alpha\beta}-\tau_{\alpha\beta}   ,
\end{equation}

\noindent
where $P$ - is the pressure, and $\tau_{\alpha\beta}$ - is 
the tensor of viscous strain, satisfying the condition of

\begin{equation}
\label{eq15}
\tau^\nu {}_\nu=0.
\end{equation}

The condition (\ref{eq15}) means that the tensor $\tau_{\alpha\beta}$ does not contain second viscosity terms. The 
fulfillment of the condition (\ref{eq15}) the representation of (\ref{eq14}) is one-valued.

It is pertinent to note that $U$ and $P$ treatment as energy density and medium pressure is in agreement with the 
treatment of the similar values in the case of energy-impulse tensor of the 
perfect liquid, i.e. in the case when

\begin{equation}
\label{eq16}
T_{\alpha\beta}=\left(U+P\right)\cdot u_\alpha u_\beta +P\cdot g_{\alpha\beta}.
\end{equation}

Here $U$ is defined by $U\equiv \left(u^\mu T_{\mu\nu} u^\nu\right)$, and $P$ is found by

\begin{equation}
\label{eq17}
P={{1}\over{3}}T_{\alpha\beta}{\left(g^{\alpha\beta}+u^\alpha u^\beta\right)}={{1}\over{3}}T_{\alpha\beta}s^{\alpha\beta}.
\end{equation}

That is in the case of the perfect liquid the formulae for 
$U$ and $P$  are the same as 
(\ref{eq13}), (\ref{eq17}) for these values in the case of CGD.

The explicit form of the values $U$, $q_\alpha$, $W_{\alpha\beta}$, $P$, $\tau_{\alpha\beta}$, 
introduced above depends on the gauge choice. When $\lambda$ constancy condition is used as 
gauge condition, Lorentz condition $A^\alpha {}_{;\alpha}=0 $ will be satisfied, and the formulae for the 
values introduced can be written in covariant form. These formulae are given 
by:

\begin{equation}
\label{eq18}
U=-{{3}\over{4}}\cdot{{\rho^2}\over{\lambda^2}}+{{1}\over{\lambda}}\left(u^\nu \rho_{;\nu}\right)-\lambda,
\end{equation}

\begin{equation}
\label{eq19}
q_\alpha=s_\alpha{}^\beta\left({{V_{;\beta}}\over{2\lambda V^2}}+{{1}\over{2\lambda V}}w_\beta\right),
\end{equation}

\begin{equation}
\label{eq20}
W_{\alpha\beta}=-{{\rho}\over{2\lambda}}\left(u_{\alpha;\beta}+u_{\beta;\alpha}\right)-{{\rho^2}\over{4\lambda^2}}\cdot s_{\alpha\beta}+
\lambda\cdot s_{\alpha\beta}+{{\rho}\over{2\lambda}}\left[u_\alpha w_\beta+u_\beta w_\alpha\right],
\end{equation}

\begin{equation}
\label{eq21}
P={{\rho^2}\over{4\lambda^2}}+\lambda+{{1}\over{3\lambda}}\left(u^\nu\rho_{;\nu}\right),
\end{equation}

\begin{equation}
\label{eq22}
\tau_{\alpha\beta}={{\rho}\over{2\lambda}} s_\alpha{}^\mu s_\beta {}^\nu\left(u_{\mu;\nu}+u_{\nu;\mu}-{{2}\over{3}}s_{\mu\nu}\left(u^\sigma{}_{;\sigma}\right)\right).
\end{equation}

The vector $w_\alpha$ from (\ref{eq19}), (\ref{eq20}) is defined by the formula $w_\alpha\equiv u^\sigma u_{\alpha ;\sigma}$, i.e. it is a four-dimensional acceleration vector.

The equations (\ref{eq18}), (\ref{eq21}) imply that there is the following connection between $U,P,V$:

\begin{equation}
\label{eq23}
P={{1}\over{3}}U+{{4}\over{3}}\lambda+{{1}\over{2V^2\lambda^2}}.
\end{equation}

These formulae is none other then the equation of state of geometrodynamic 
continuum. 

The formula for isentropic sound speed $c_s$, defined as

\begin{equation}
\label{eq24}
c_s^2 =-V^2\left({{\partial P}\over{\partial V}}\right),
\end{equation}

\noindent
is given by

\begin{equation}
\label{eq25}
c_s^2={{4}\over{3}}V\left(P-\lambda\right)+{{1}\over{2\lambda^2 V}}.
\end{equation}

\section{Equations for vector and antisymmetric tensor in CGD in the case of flat 
space}
\label{sec:equations}

Equation (\ref{eq1}) with the right side in the form of tensor (\ref{eq2}) implies that 
formulae

\begin{equation}
\label{eq26}
T_\alpha {}^\beta {} _{;\beta } = 0
\end{equation}

\noindent
should be performed. When the antisymmetric tensor is introduced

\begin{equation}
\label{eq27}
F_{\alpha \beta } = A_{\beta ,\alpha } - A_{\alpha ,\beta }, 
\end{equation}

\noindent
the formulae

\begin{equation}
\label{eq28}
F_{\alpha \cdot } {}^\beta  {}_{;\beta } = \lambda _{;\alpha } - 2\lambda 
A_\alpha .
\end{equation}

\noindent
follows from (\ref{eq26}). If

\begin{equation}
\label{eq29}
\lambda = \mbox{Const},
\end{equation}

\noindent
is taken as gauge condition, then Weyl vector will satisfy Lorentz condition

\begin{equation}
\label{eq30}
A^\nu  {}_{;\nu } = 0.
\end{equation}

There are problems in CGD and GTR in which the evolution of gravitational 
degrees of freedom as well as degrees of freedom connected with matter 
fields should be consistently considered. Among these problems is 
Schwarzchild problem. On the other hand there is a large number of problems 
in physics which are set up to consider the matter fields dynamics without 
regard of gravitational degree of freedom. Here the space is set up to be 
flat, while the dynamic equations for matter fields are derived from (\ref{eq26}). 

The flat space problem set up in the case of CGD equations implies that the 
dynamic equations should result from (\ref{eq26}) for the complementary degree of 
freedom providing the energy-momentum tensor used for tensor (\ref{eq2}). In case 
that $T_\alpha^\beta{}_{;\beta}$ is 
obtained, the equation is changed with (\ref{eq1}), and given the gauge (\ref{eq29}), we get 
the following four equations:

\begin{equation}
\label{eq31}
J_{\beta ,\alpha } - J_{\alpha ,\beta } = 4m\cdot H_{\alpha \beta } ,
\end{equation}

\begin{equation}
\label{eq32}
H_\alpha {}^\beta {}_{;\beta } = -m\cdot J_\alpha,
\end{equation}

\begin{equation}
\label{eq33}
J^\nu _{;\nu } = 0 
\end{equation}

\begin{equation}
\label{eq34}
m=Const.
\end{equation}

Providing the initial CGD equations written in terms of Weyl vector 
$A_\alpha$ and lambda term $\lambda$, the equations 
(\ref{eq31})-(\ref{eq34}) are written in new terms: $J^\alpha, \ H^{\alpha\beta}, \ m$. The connection
between the  values depends on the 
gauge conditions. Here

\begin{equation}
\label{eq35}
J_\alpha={{2}\over{m}}A_\alpha; \, H_{\alpha\beta}={{1}\over{2m^2}}F_{\alpha\beta},
\end{equation}

\begin{equation}
\label{eq36}
\lambda=2m^2.
\end{equation}

\section{Dirac matrices. Dirac equation}
\label{sec:dirac}

For the coherency of the treatment and convenience of the references let us 
mention some properties of Dirac matrices and Dirac equation which will be 
used later.

Our concern will be the case when the space can be considered plane. In this 
case the metric tensor of Riemannian space $g_{\alpha\beta}$ 
can be considered the same as metric tensor of 
Minkowsky space, which is given by $g_{\alpha\beta}=diag\left[-1,1,1,1\right]$
in Cartesian coordinates. Dirac matrices (DM) 
$\gamma_\alpha$ are defined by

\begin{equation}
\label{eq37}
\gamma_\alpha \gamma_\beta+\gamma_\beta \gamma_\alpha=2g_{\alpha\beta}\cdot E.
\end{equation}

The symbol $E$ means the identity matrix $4\times 4$ 
in (\ref{eq37}). DM $\gamma_\alpha$ are 
constant in the entire space. We use Majorana system of matrices

\begin{equation}
\label{eq38}
\gamma_0=-i\rho_2\sigma_1, \, \gamma_1=\rho_1, \, \gamma_2=\rho_2\sigma_2, \, 
\gamma_3=\rho_3,
\end{equation}

\noindent
in case the explicit form of DM is required. The elements of this system are 
integral real numbers\footnote{ There is a real system of DM among (\ref{eq37}) 
solutions in case the signature $\left(-+++\right)$ is used. }. 

Suppose that field functional $Z$ is general the $4\times 4$ matrix satisfying Dirac equation

\begin{equation}
\label{eq39}
\gamma^\nu\left(\nabla_nu Z\right)=m\cdot Z.
\end{equation}

By $Z $ we shall 
mean the bispinor matrix. The bispinor states are selected from 
$Z$ by its right multiplication by the projections.

The equation 

\begin{equation}
\label{eq40}
\left(\nabla_\nu Z^{+}\right)D\gamma^\nu=-m\cdot Z^{+}D.
\end{equation}

\noindent
is combined with (\ref{eq39}). $D$ matrix in (\ref{eq40}) is defined by

\begin{equation}
\label{eq41}
D\gamma_\mu D^{-1}=-\gamma_\mu^{+}.
\end{equation}

\noindent
Covariant derivatives of bispinor matrix in (\ref{eq39}), (\ref{eq40}), are set down as:

\begin{equation}
\label{eq42}
\left. {\begin{array}{l}
 \nabla _\alpha Z = Z_{;\alpha } - Z\Gamma _\alpha \\ 
 \nabla _\alpha Z^ + = Z^ + _{;\alpha } + \Gamma _\alpha Z^+ \\ 
 \end{array}} \right\} .
\end{equation}
\noindent
The value of $\Gamma_\alpha$ in 
(\ref{eq42}) will be referred to as bispinor connectivity\footnote{ The bispinor 
connectivity and gauge field agree within constant factor.}. It is the whole 
complex of real anti-Hermitean matrices:

\begin{equation}
\label{eq43}
\Gamma ^\ast _\alpha = \Gamma _\alpha ,\quad \Gamma ^ + _\alpha = - \Gamma_\alpha .
\end{equation}

Let us mention some formulae resulting from Dirac equation. Equation (\ref{eq39}) is 
multiplied by $\gamma^\alpha$ on the left side.

\begin{equation}
\label{eq44}
\gamma^\alpha \gamma^\nu\left(\nabla_\nu Z\right)=m\cdot\gamma^\alpha Z .
\end{equation}

\noindent
Let us write the product $\gamma^\alpha \gamma^\nu$ as

\begin{equation}
\label{eq45}
\gamma^\alpha \gamma^\nu=g^{\alpha\nu}+S^{\alpha\nu}
\end{equation} 
\noindent
(here $S^{\mu\nu}={{1}\over{2}}\left(\gamma_\mu \gamma_\nu-\gamma_\nu \gamma_\mu\right)$ 
and (\ref{eq45}) are inserted into (\ref{eq44}).

\begin{equation}
\label{eq46}
\left( {\nabla_\alpha Z} \right) = -  {S}_\alpha {}^\nu \left( {\nabla 
_\nu Z} \right) + m \cdot {\gamma }_\alpha Z.
\end{equation}
\noindent
After Hermitean conjugation (\ref{eq46}) and $D$ 
multiplication from the right side we get:

\begin{equation}
\label{eq47}
\left( {\nabla _\alpha Z^ + } \right) {D} = \left( {\nabla _\nu Z^ + } 
\right) {D} {S}_\alpha {}^\nu - m \cdot Z^ + {D} 
{\gamma }_\alpha .
\end{equation}

\section{Tensor mapping onto bispinor matrix}
\label{sec:tensor}

Suppose that the set of five types of tensors enumerated in 
\ref{tab1} is defined in 4D Riemannian space. 

Let us construct the matrix $M$ according to the rule

\begin{equation}
\label{eq48}
M\equiv a\cdot iD^{-1}+b\cdot i\gamma_5 D^{-1}+J_\alpha\cdot\gamma^\alpha D^{-1}
+s_\alpha\cdot i\gamma_5\gamma^\alpha D^{-1}+H_{\alpha\beta}\cdot S^{\alpha\beta} D^{-1}.
\end{equation}

Here $\gamma_5=\gamma_0 \gamma_1 \gamma_2 \gamma_3$. Each of the 
summands in the right side of (\ref{eq48}) is Hermitian matrix, so that 
$M$ matrix is Hermitian as well. 
The full number of tensor components enumerated in Table \ref{tab1}, is 16.

\begin{table}[htbp]
\begin{tabular}
{|p{20pt}|p{151pt}|p{78pt}|}
\hline
& 
Tensor type& 
Symbol \\
\hline
1& Scalar& $a$ \\
\hline
2& Vector& $J^\alpha$ \\
\hline
3& Antisymmetric tensor& $H^{\alpha\beta}$ \\
\hline
4& Pseudovector& $s^\alpha$ \\
\hline
5& Pseudoscalar& $b$ \\
\hline
\end{tabular}
\caption{-- Five types of tensor}
\label{tab1}
\end{table}

\noindent
$M$ matrix is changed as $M\rightarrow M^{\prime}=LML^{+}$ under Lorentz 
transformations of DM $\gamma_\alpha\rightarrow\gamma^{\prime}{}_\alpha=L\gamma_\alpha L^{-1}$, i.e. the matrix is the object of bispinor product type by 
conjugated Hermitian bispinor.

\begin{equation}
\label{eq49}
If\hskip5mm \gamma_\alpha\rightarrow\gamma^{\prime}{}_\alpha=L\gamma_\alpha L^{-1},
\hskip5mm then\hskip5mm  M\rightarrow M^{\prime}=LML^{+}.
\end{equation}

Being Hermitian, the matrix (\ref{eq48}) can have any rank up to four included. When 
the rank is given, the eigenvalue spectrum can include the real values 
(positive and negative), as well as complex values (complex conjugated 
values). Here we confine ourselves to the analysis of the space-time regions 
in which the rank of matrix (\ref{eq48}) is 4 and all the eigenvalues are positive. 

It is known from the general theory of matrices that ``the square root'' can 
be taken from $M$ i.e. $M$ matrix can be represented as

\begin{equation}
\label{eq50}
M=Z\cdot Z^{+}.
\end{equation}

\noindent
If $Z$ in (\ref{eq50}) implies 
the arithmetical root, then the procedure of ``square-root'' generation is 
single-valued. 

(\ref{eq48}) and (\ref{eq50}) imply that the original tensor are related to 
$Z$ with

\begin{equation}
\label{eq51}
\left.
\begin{array}{c}
a=-{{i}\over{4}}\cdot {\Large Sp}\left(Z^{+}DZ\right),\hskip5mm 
b={{i}\over{4}}\cdot {\Large Sp}\left(Z^{+}D\gamma_5 Z\right),\\ 
s^\alpha=-{{i}\over{4}}\cdot {\Large Sp}\left(Z^{+}D\gamma_5 \gamma^\alpha Z\right),\\
J^\alpha={{1}\over{4}}\cdot {\Large Sp}\left(Z^{+}D \gamma^\alpha Z\right),\hskip5mm
H^{\alpha\beta}=-{{1}\over{8}}\cdot {\Large Sp}\left(Z^{+}DS^{\alpha\beta} Z\right).
\end{array}\right\}
\end{equation}

The fulfillment of the condition of tractability (\ref{eq50}) means that

\begin{equation}
\label{eq52}
det\left( Z\right)\neq 0
\end{equation}

\noindent
as well as that there is $Z^{-1}$ matrix along with $Z$ matrix.

In particular, the real numbers are quite sufficient to solve the problem of 
tensor mapping onto the bispinor matrix when only $J^\alpha$ and $H^{\alpha\beta}$ tensor are nonzero among all the 
tensors enumerated in the Table \ref{tab1} , and DM are used 
as in real representation as DM system. This case will be of the particular 
concern. The full number of $J^\alpha$ and $H^{\alpha\beta}$ components is 10. In this particular case Hermitian 
$M$ matrix is given by

\begin{equation}
\label{eq53}
M=J^\alpha\cdot\left(\gamma_\alpha D^{-1}\right)+
H^{\alpha\beta}\cdot\left(S_{\alpha\beta} D^{-1}\right).
\end{equation}

All the mapping results presented above have the algebraic character as they 
belong to tensor and bispinor matrices at one arbitrary point of Riemannian 
space. It stands to reason that if the fields $J^\alpha\left(x\right),\, 
H^{\alpha\beta}\left(x\right)$ are defined in certain region of space and if the condition of 
$M\left(x\right)$ matrix 
positivity is fulfilled, then $Z\left(x\right)$ matrix is set up at each point and hence we have the mapping 
onto the bispinor matrix of two tensor fields: $J^\alpha\left(x\right),\, 
H^{\alpha\beta}\left(x\right)$. 

\section{Dirac equation in terms of the observed values}
\label{sec:mylabel2}

Suppose that the fields $J^\alpha\left(x\right),\, 
H^{\alpha\beta}\left(x\right)$ are set 
in an arbitrary region of space and at each point $M\left(x\right)$ matrix can be mapped onto 
$M\left(x\right)$ bispinor 
matrix. Let us raise the question: from which law will $Z$ matrix vary if the vector and 
antisymmetric vector from (\ref{eq53}) obey CGD equations, i.e. (\ref{eq31})-(\ref{eq34})?

The answer can be obtained from three theorems prooof given below.

\subsection{Theorem 1}
\label{subsec:theorem}

Let us put Theorem 1 as

\begin{equation}
\label{eq54}
J^\alpha {}_{;\alpha}=0.
\end{equation}

Let us prove that first, it is fulfilled if the vector $J^\alpha$ is expressed in terms of bispinor 
matrix according to the formula (\ref{eq51}),

\[
J^\alpha={{1}\over{4}}\cdot {\Large Sp}\left(Z^{+}D \gamma^\alpha Z\right),
\]

\noindent
and second, $Z, \, Z^{+}$ 
matrices satisfy Dirac equations as (\ref{eq39}) and (\ref{eq40}).
Let us differentiate the formula for $J^\alpha$.

\[
J^\alpha {}_{;\alpha}={{1}\over{4}}\cdot {\Large Sp}\left(\left(\nabla_\alpha Z^{+}\right)D \gamma^\alpha Z\right)+
{{1}\over{4}}\cdot {\Large Sp}\left(Z^{+}D \gamma^\alpha \left(\nabla_\alpha Z\right)\right).
\]
\noindent
We use Dirac equation.

\[
J^\alpha {}_{;\alpha}={{1}\over{4}}\cdot {\Large Sp}\left(-m\cdot Z^{+}DZ+m\cdot Z^{+}DZ\right)=0.
\]

Hence, the theorem is proved.

\subsection{Theorem 2}
\label{subsec:mylabel2}

Theorem 2 states that if bispinor matrix obeys Dirac equation (\ref{eq39}) and the 
values of $J_\alpha$, $H_{\alpha\beta}$ are related with 
(\ref{eq51}), then the formula 

\begin{equation}
\label{eq55}
\begin{array}{c}
\left( J_{\beta ,\alpha } - J_{\alpha ,\beta }  \right) = 4m \cdot 
H_{\alpha \beta } +\\
+ E_{\alpha \beta \mu }  {}^\nu \frac{1}{4}\mbox{Sp}\left\{ 
{\left( {\nabla _\nu Z^ + } \right){D}{\gamma }_5  
{\gamma }^\mu Z - Z^ + {D} {\gamma }_5  {\gamma }^\mu 
\left( {\nabla _\nu Z} \right)} \right\}
\end{array}
\end{equation}

\noindent
is true. 

The theorem will be proved in several steps. First, we try to value 
$\left( {J_{\beta ,\alpha } - J_{\alpha ,\beta } } \right) $, using (\ref{eq46}), (\ref{eq47}). (\ref{eq51}) implies:

\begin{equation}
\label{eq56}
\begin{array}{l}
 \left( {J_{\beta ,\alpha } - J_{\alpha ,\beta } } \right) = 
\noindent
\frac{1}{4}\mbox{Sp}\left\{ {\left( {\nabla _\alpha Z^ + } \right) 
{D} {\gamma }_\beta Z + Z^ +  {D} {\gamma }_\beta \left( 
{\nabla _\alpha Z} \right)} \right\} - \\ 
 - \frac{1}{4}\mbox{Sp}\left\{ {\left( {\nabla _\beta Z^ + } \right) 
{D} {\gamma }_\alpha Z + Z^ +  {D} {\gamma }_\alpha \left( 
{\nabla _\beta Z} \right)} \right\}. \\ 
 \end{array}
\end{equation}

Use (\ref{eq46}), (\ref{eq47}).

\begin{equation}
\label{eq57}
\begin{array}{l}
 \left( {J_{\beta ,\alpha } - J_{\alpha ,\beta } } \right) = \\ 
 = \frac{1}{4}\mbox{Sp}\left\{ \left( {\nabla_\nu Z^ + } \right) 
{D} {S}_\alpha  {}^\nu  {\gamma }_\beta Z - m \cdot Z^ + 
{D} {\gamma }_\alpha  {\gamma }_\beta Z - Z^+  {D} 
{\gamma }_\beta  {S}_\alpha  {}^\nu \left( {\nabla _\nu Z} 
\right) + \right.  \\
\left. + m \cdot Z^ +  {D} {\gamma }_\beta  
{\gamma }_\alpha Z 
\right\} 
 - \frac{1}{4}\mbox{Sp}\left\{ \left( {\nabla _\nu Z^+ } \right) 
{D} {S}_\beta {}^\nu  {\gamma }_\alpha Z - \right. \\
\left. -m \cdot Z^ + {D} {\gamma }_\beta  {\gamma }_\alpha Z - Z^ +  {D} 
{\gamma }_\alpha  {S}_\beta  {}^\nu \left( {\nabla _\nu Z} \right. + m 
\cdot Z^ +  {D} {\gamma }_\alpha  {\gamma }_\beta Z 
\right\}. \\ 
 \end{array}
\end{equation}
\noindent
Let us combine the terms with and without the derivatives in the right side 
of (\ref{eq57}) separately.

\begin{equation}
\label{eq58}
\begin{array}{l}
 \left( {J_{\beta ,\alpha } - J_{\alpha ,\beta } } \right) = \frac{1}{2}m 
\cdot \mbox{Sp}\left\{ { - Z^ +  {D} {\gamma }_\alpha  
{\gamma }_\beta Z + Z^ +  {D}{\gamma }_\beta  {\gamma 
}_\alpha Z} \right\} \\ 
 + \frac{1}{4}\mbox{Sp}\left\{ {\left( {\nabla _\nu Z^ + } \right) 
{D} {S}_\alpha ^\nu  {\gamma }_\beta Z - Z^ +  {D} 
{\gamma }_\beta  {S}_\alpha  {}^\nu \left( {\nabla _\nu Z} \right) - 
\left( {\nabla _\nu Z^ + } \right) {D} {S}_\beta {}^\nu  
{\gamma }_\alpha Z }\right.\\
\left. + Z^+  {D} {\gamma }_\alpha  {S}_\beta  {}^\nu \left( {\nabla _\nu Z} \right) \right\}. \\ 
 \end{array}
\end{equation}
\noindent
Let us replace DM product such as $ {S}_\alpha  {}^\nu  {\gamma }_\beta $, 
$ {\gamma }_\beta  {S}_\alpha ^\nu $, in (\ref{eq58}) using the following formulae:

\[
\left. {\begin{array}{l}
 {S}_\alpha  {}^\nu  {\gamma }_\beta = - \eta _{\alpha \beta } 
 {\gamma }^\nu + \delta _\beta ^\nu  {\gamma }_\alpha + E_\alpha  {}^\nu  {}_{\beta \mu }  {\gamma }_5  {\gamma }^\mu \\ 
  {\gamma }_\beta  {S}_\alpha  {}^\nu = \eta _{\alpha \beta } 
 {\gamma }^\nu - \delta _\beta ^\nu  {\gamma }_\alpha + E_\alpha 
{}^\nu  {}_{\beta \mu }  {\gamma }_5  {\gamma }^\mu \\ 
 \end{array}} \right\} .
\]

\noindent
We get:

\begin{equation}
\label{eq59}
\begin{array}{l}
 \left( {J_{\beta ,\alpha } - J_{\alpha ,\beta } } \right) = - m \cdot 
\mbox{Sp}\left\{ {Z^ +  {D} {S}_{\alpha \beta } Z} \right\} \\ 
 + \frac{1}{4}\mbox{Sp}\left\{ {\left( {\nabla _\nu Z^ + } \right) 
{D}\left( { - \eta _{\alpha \beta }  {\gamma }^\nu + \delta _\beta 
{}^\nu  {\gamma }_\alpha + E_{\alpha} {}^{\nu} {}_{\beta \mu }  {\gamma 
}_5  {\gamma }^\mu } \right)Z} \right\} \\ 
 + \frac{1}{4}\mbox{Sp}\left\{ { - Z^ +  {D}\left( {\eta _{\alpha 
\beta }  {\gamma }^\nu - \delta _\beta ^\nu  {\gamma }_\alpha + 
E_\alpha {}^\nu  {}_{\beta \mu }  {\gamma }_5  {\gamma }^\mu } 
\right)\left( {\nabla _\nu Z} \right)} \right\} \\ 
 + \frac{1}{4}\mbox{Sp}\left\{ { - \left( {\nabla _\nu Z^ + } \right) 
{D}\left( { - \eta _{\alpha \beta }  {\gamma }^\nu + \delta _\alpha 
{}^\nu  {\gamma }_\beta - E_\alpha {}^\nu  {}_{\beta \mu }  {\gamma }_5 
 {\gamma }^\mu } \right)Z} \right\} \\ 
 + \frac{1}{4}\mbox{Sp}\left\{ { + Z^ +  {D}\left( {\eta _{\alpha 
\beta }  {\gamma }^\nu - \delta _\alpha ^\nu  {\gamma }_\beta - 
E_\alpha {}^\nu  {}_{\beta \mu }  {\gamma }_5  {\gamma }^\mu } 
\right)\left( {\nabla _\nu Z} \right)} \right\}. \\ 
 \end{array}
\end{equation}
\noindent
The terms in (\ref{eq59}), containing the metric tensor, are reduced. The other 
give:

\[
\begin{array}{l}
 \left( {j_{\beta ,\alpha } - j_{\alpha ,\beta } } \right) = + 8m \cdot 
h_{\alpha \beta } \\ 
 + \frac{1}{4}\mbox{Sp}\left\{ {\left( {\nabla _\beta Z^ + } \right) 
{D} {\gamma }_\alpha Z} \right\} + E_\alpha {}^\nu {}_{\beta \mu } 
\frac{1}{4}\mbox{Sp}\left\{ {\left( {\nabla _\nu Z^ + } \right) 
{D} {\gamma }_5  {\gamma }^\mu Z} \right\} \\ 
 + \frac{1}{4}\mbox{Sp}\left\{ {Z^ +  {D} {\gamma }_\alpha 
\left( {\nabla _\beta Z} \right)} \right\} - E_\alpha {}^\nu  {}_{\beta \mu } 
\frac{1}{4}\mbox{Sp}\left\{ {Z^ +  {D} {\gamma }_5  
{\gamma }^\mu \left( {\nabla _\nu Z} \right)} \right\} \\ 
 + \frac{1}{4}\mbox{Sp}\left\{ { - \left( {\nabla _\alpha Z^ + } 
\right) {D} {\gamma }_\beta Z} \right\} + E_\alpha {}^\nu {}_{\beta 
\mu } \frac{1}{4}\mbox{Sp}\left\{ {\left( {\nabla _\nu Z^ + } \right) 
{D} {\gamma }_5  {\gamma }^\mu Z} \right\} \\ 
 + \frac{1}{4}\mbox{Sp}\left\{ { - Z^ +  {D} {\gamma }_\beta 
\left( {\nabla _\alpha Z} \right)} \right\} - E_\alpha {}^\nu {}_{\beta \mu } 
\frac{1}{4}\mbox{Sp}\left\{ { + Z^ +  {D} {\gamma }_5  
{\gamma }^\mu \left( {\nabla _\nu Z} \right)} \right\}. \\ 
 \end{array}
\]

The terms containing $D\gamma_\alpha$, are reduced to 
$-\left(J_{\beta,\alpha}-J_{\alpha,\beta}\right)$. The other are combined into 
$E_\alpha {}^\nu  {}_{\beta \mu } \frac{1}{2}\mbox{Sp}\left\{ {\left( {\nabla _\nu Z^ + } 
\right) {D} {\gamma }_5  {\gamma }^\mu Z - Z^ +  
{D} {\gamma }_5  {\gamma }^\mu \left( {\nabla _\nu Z} \right)} 
\right\}$. As a result we get

\begin{equation}
\label{eq60}
\begin{array}{l}
\left( {J_{\beta ,\alpha } - J_{\alpha ,\beta } } \right) = 8m \cdot 
H_{\alpha \beta } - \left( {J_{\beta ,\alpha } - J_{\alpha ,\beta } } 
\right) \\
+ E_\alpha {}^\nu  {}_{\beta \mu } \frac{1}{2}\mbox{Sp}\left\{ {\left( 
{\nabla _\nu Z^ + } \right) {D} {\gamma }_5  {\gamma }^\mu 
Z - Z^ +  {D} {\gamma }_5  {\gamma }^\mu \left( {\nabla 
_\nu Z} \right)} \right\}.\\
\end{array}
\end{equation}

We get the formula which agrees with (\ref{eq55}) after identity transformations of 
(\ref{eq60}). Hence, the theorem 2 is proved.

\subsection{Theorem 3}
\label{subsec:mylabel3}

Theorem 3 states that the formula

\begin{equation}
\label{eq61}
H_\alpha {}^\nu  {}_{;\nu } = - \frac{1}{8}\mbox{Sp}\left( {\left( {\nabla 
_\alpha Z^ + } \right) {D}Z - Z^+  {D}\left( {\nabla _\alpha Z} 
\right)} \right) - m \cdot J_\alpha 
\end{equation}

\noindent
is true. 

The proof involves the direct check of validity of (\ref{eq61}). We have:

\begin{equation}
\label{eq62}
\begin{array}{l}
 H_\alpha {}^\nu  {}_{;\nu } = - \frac{1}{8}\mbox{Sp}\left\{ {Z^ +  
{D} {S}_\alpha  {}^\nu Z} \right\}_{;\nu } = \\ 
 = - \frac{1}{8}\mbox{Sp}\left\{ {\left( {\nabla _\nu Z^ + } \right) 
{D} {S}_\alpha  {}^\nu Z} \right\} - \frac{1}{8}\mbox{Sp}\left\{ {Z^ + 
 {D} {S}_\alpha  {}^\nu \left( {\nabla _\nu Z} \right)} \right\}. 
\\ 
 \end{array}
\end{equation}

\noindent
In the first case we change the matrix $S_\alpha {}^\nu$ according to

\begin{equation}
\label{eq63}
 {S}_\alpha  {}^\nu = \delta _\alpha ^\nu -  {\gamma }^\nu {\gamma }_\alpha ,
\end{equation}

\noindent
and in the second case according to

\begin{equation}
\label{eq64}
 {S}_\alpha  {}^\nu = - \delta _\alpha ^\nu +  {\gamma }_\alpha  {\gamma }^\nu .
\end{equation}

\noindent
We get:

\begin{equation}
\label{eq65}
\begin{array}{l}
 H_\alpha {}^\nu  {}_{;\nu } = - \frac{1}{8}\mbox{Sp}\left\{ {\left( {\nabla _\nu 
Z^ + } \right) {D}\left( {\delta _\alpha ^\nu -  {\gamma }^\nu 
 {\gamma }_\alpha } \right)Z} \right\} \\
- \frac{1}{8}\mbox{Sp}\left\{ 
{Z^+  {D}\left( { - \delta _\alpha ^\nu +  {\gamma }_\alpha 
 {\gamma }^\nu } \right)\left( {\nabla _\nu Z} \right)} \right\} = \\ 
 = - \frac{1}{8}\mbox{Sp}\left\{ {\left( {\nabla _\nu Z^ + } \right) 
{D}\left( {\delta _\alpha ^\nu } \right)Z} \right\} - 
\frac{1}{8}\mbox{Sp}\left\{ {\left( {\nabla _\nu Z^ + } \right) 
{D}\left( { -  {\gamma }^\nu  {\gamma }_\alpha } \right)Z} 
\right\} \\ 
 - \frac{1}{8}\mbox{Sp}\left\{ {Z^ +  {D}\left( { - \delta _\alpha 
^\nu } \right)\left( {\nabla _\nu Z} \right)} \right\} - 
\frac{1}{8}\mbox{Sp}\left\{ {Z^ +  {D}\left( { {\gamma }_\alpha 
 {\gamma }^\nu } \right)\left( {\nabla _\nu Z} \right)} \right\} = \\ 
 = - \frac{1}{8}\mbox{Sp}\left\{ {\left( {\nabla _\alpha Z^ + } 
\right) {D}Z - Z^ +  {D}\left( {\nabla _\alpha Z} \right)} 
\right\} \\ 
 + \frac{1}{8}\mbox{Sp}\left\{ {\left( {\nabla _\nu Z^ + } \right) 
{D} {\gamma }^\nu  {\gamma }_\alpha Z} \right\} - 
\frac{1}{8}\mbox{Sp}\left\{ {Z^ +  {D} {\gamma }_\alpha  
{\gamma }^\nu \left( {\nabla _\nu Z} \right)} \right\} \\ 
 \end{array}
\end{equation}

\noindent
After using (\ref{eq39}), (\ref{eq40}) we get:

\begin{equation}
\label{eq66}
H_\alpha {}^\nu  {}_{;\nu } = \frac{1}{8}\mbox{Sp}\left\{ {\left( {\nabla _\alpha 
Z^ + } \right) {D}Z - Z^ +  {D}\left( {\nabla _\alpha Z} 
\right)} \right\} - 2m \cdot \frac{1}{8}\mbox{Sp}\left\{ {Z^ +  
{D} {\gamma }_\alpha Z} \right\}.
\end{equation}

The obtained relation (\ref{eq66}) agrees with (\ref{eq61}). Hence, it is proved that the 
relation (\ref{eq61}) follows from Dirac equation, i.e. the theorem 3 is proved. 

The results of the proved theorems are summed in Table \ref{tab2}.

\begin{table}
\label{tab2}
\begin{tabular}
{|p{40pt}|p{301pt}|}
\hline
Theorem number& Theorem tells that\\
\hline
1&$J^\alpha {}_{;\alpha}=0$ \\
\hline
2& 
$\left( {J_{\beta ,\alpha } - J_{\alpha ,\beta } } \right) = 4m \cdot 
H_{\alpha \beta } +$ \\
&$+E_{\alpha \beta \mu }  {}^\nu \frac{1}{4}\mbox{Sp}\left\{ 
{\left( {\nabla _\nu Z^ + } \right){D}{\gamma }_5  
{\gamma }^\mu Z - Z^ + {D} {\gamma }_5  {\gamma }^\mu 
\left( {\nabla _\nu Z} \right)} \right\}$ \\
\hline
3& $H_\alpha {}^\nu _{;\nu } = - \frac{1}{8}\mbox{Sp}\left( {\left( {\nabla 
_\alpha Z^ + } \right) {D}Z - Z^+  {D}\left( {\nabla _\alpha Z} 
\right)} \right) - m \cdot J_\alpha $ \\
\hline
\end{tabular}
\caption{-- Relations resulting from Dirac equation}
\label{tab2}
\end{table}

If all the spur terms in Table \ref{tab2} are zero, the 
relation between the vector $J^\alpha$ and the tensor $H^{\alpha\beta}$ 
will take the form which agree completely with (\ref{eq31})-(\ref{eq34}). Let us 
prove that it is necessary to set bispinor connectivity in 
Table \ref{tab2} as

\begin{equation}
\label{eq67}
\Gamma_\alpha={{1}\over{2}}\left[ \left(Z^{-1}Z_{;\alpha}\right)-
\left(Z^{+}_{;\alpha}Z^{-1}{}^{+}\right)\right]
\end{equation}

\noindent
to make the spur terms zero. Actually, it is necessary to prove that on 
substituting (\ref{eq67}) in spur terms, these terms go to zero, i.e. the following 
equalities are performed:

\begin{equation}
\label{eq68}
E_{\alpha\beta\mu}{}^\nu { Sp}\left\{\left(\nabla_\nu Z^{+}\right)D\gamma_5 \gamma^\mu Z\right. -
\left. Z^{+} D\gamma_5 \gamma^\mu \left(\nabla_\nu Z\right)\right\}=0.
\end{equation}

\begin{equation}
\label{eq69}
{ Sp}\left\{\left(\nabla_\alpha Z^{+}\right)D Z\right. -
\left. Z^{+} D \left(\nabla_\alpha Z\right)\right\}=0.
\end{equation}

Let us check it by the example of one of the equalities (\ref{eq68}) - (\ref{eq69}). For 
example, the equality (\ref{eq69}). We use spur properties.

\[
\begin{array}{l}
{ Sp}\left\{\left(\nabla_\alpha Z^{+}\right)D Z\right. -
\left. Z^{+} D \left(\nabla_\alpha Z\right)\right\}=\\
=2\cdot { Sp}\left[ \Gamma_\alpha \left(Z^{+} DZ\right)\right]+
{ Sp}\left\{Z^{+}_{;\alpha}DZ-Z^{+}DZ_{;\alpha}\right\}\\
\end{array}
\]

\[
\begin{array}{l}
{ Sp}\left\{\left(\nabla_\alpha Z^{+}\right)D Z\right. -
\left. Z^{+} D \left(\nabla_\alpha Z\right)\right\}=\\
=2\cdot { Sp}\left[ \Gamma_\alpha \left(Z^{+} DZ\right)\right]+
{ Sp}\left\{Z^{+}_{;\alpha}DZ-Z^{+}DZ_{;\alpha}\right\}\\
\end{array}
\]

The identity transformations show that

\[
Sp\left[\left(\nabla_\alpha Z^{+}\right)DZ-Z^{+}D\left(\nabla_\alpha Z\right)\right]=0,
\]

\noindent
i.e. the equality (\ref{eq69}) is fulfilled when bispinor connectivity is defined by 
(\ref{eq67}). The validity of (\ref{eq68}) is proved in a similar way.

\section{The example of the exact solution of CGD equations }
\label{sec:mylabel1}

\subsection{The exact solution}
\label{subsec:mylabel1}

The solution concerned takes the form:

\begin{equation}
\label{eq70}
\begin{array}{l}
J_0=u,\hskip5mm J_k=0\\
H_{0k}=-{{1}\over{4m}}u' {{x_k}\over{r}},\hskip5mm H_{mn}=0.\\
\end{array}
\end{equation}

\noindent
Here

\begin{equation}
\label{eq71}
u=u(r),
\end{equation}

\noindent
so the solution is steady-state and spherically symmetrical. The Anzats (\ref{eq70}) 
provides automatic execution of the equations (\ref{eq31}), (\ref{eq33}), (\ref{eq34}). The 
equations (\ref{eq32}) should be satisfied in order that (\ref{eq70}) is the solution of CGD 
equations. Substitute (\ref{eq70}) in (\ref{eq32}). It turns out that (\ref{eq32}) is satisfied if 
$u$ function is the solution of

\begin{equation}
\label{eq72}
u''+{{2}\over{r}}u'-4m^2 u=0.
\end{equation}

The general solution of (\ref{eq72}) consists of two summands:

\begin{equation}
\label{eq73}
u=C_1 \cdot{{e^{-2mr}}\over{mr}}+C_2 \cdot{{e^{2mr}}\over{mr}}.
\end{equation}

\noindent
Each summand is included in (\ref{eq73}) with dimensionless integration 
$C_1, C_2$. We shall consider 
the case when $C_1=0$. If $m$ constant is 
positive, the solution of (\ref{eq73}) grows exponentially,

\begin{equation}
\label{eq74}
u=-c\cdot{{e^{2mr}}\over{mr}}.
\end{equation}

\noindent
In (\ref{eq74}) $C_2$ constant stands for $-c$. Let us substitute (\ref{eq74}) in (\ref{eq70}) and get:

\begin{equation}
\label{eq75}
J_0=-c\cdot {{e^{2mr}}\over{mr}},\hskip5mm H_{0k}=-{{c}\over{4m^2}}\cdot e^{2mr}\cdot\left[{{12}\over{r^2}}-{{2m}\over{r}}\right]
\cdot{{x_k}\over{r}}.
\end{equation}

Let us substitute (\ref{eq75}) in (\ref{eq53}) and use DM in the form of (\ref{eq38}) to obtain 
$M$ matrix. We get:

\begin{equation}
\label{eq76}
M=
\begin{array}{|c|c|c|c|}\hline
-u-{{1}\over{2m}}u'{{z}\over{r}}&0&-{{1}\over{2m}}u'{{x}\over{r}}&{{1}\over{2m}}u'{{y}\over{r}}\\ \hline
0&-u-{{1}\over{2m}}u'{{z}\over{r}}&-{{1}\over{2m}}u'{{y}\over{r}}&-{{1}\over{2m}}u'{{x}\over{r}}\\ \hline
-{{1}\over{2m}}u'{{x}\over{r}}&-{{1}\over{2m}}u'{{y}\over{r}}&-u'+{{1}\over{2m}}u'{{z}\over{r}}&0\\ \hline
{{1}\over{2m}}u'{{y}\over{r}}&-{{1}\over{2m}}u'{{x}\over{r}}&0&-u'+{{1}\over{2m}}u'{{z}\over{r}}\\ \hline
\end{array}
\end{equation}
\noindent
The eigenvalues of $M$ matrix:

\begin{equation}
\label{eq77}
\mu_1=-u-{{1}\over{2m}}u';\hskip2mm \mu_2=-u-{{1}\over{2m}}u';\hskip2mm \mu_3=-u+{{1}\over{2m}}u';\hskip2mm \mu_4=-u+{{1}\over{2m}}u'.
\end{equation}

\noindent
The substitution of (\ref{eq74}) in (\ref{eq77}) for $u$ gives:

\begin{equation}
\label{eq78}
\mu_1=\mu_2=-{{c\cdot e^{2mr}}\over{2m^2 r^2}}\cdot\left[1-4mr\right];\hskip5mm \mu_3=\mu_4=-{{c\cdot e^{2mr}}\over{2m^2 r^2}}.
\end{equation}

\noindent
It follows from (\ref{eq78}) that positivity condition of all eigenvalues first 
consists of constant $c$ constraint

\begin{equation}
\label{eq79}
c>0,
\end{equation}

\noindent
second, of constraints of the range of radial variable values

\begin{equation}
\label{eq80}
r>1/4m.
\end{equation}

\noindent
Some of the eigenvalues are negative when the conditions (\ref{eq79}), (\ref{eq80}) are 
violated.

Let us denote the normalized eigenvectors corresponding to the eigenvalues 
(\ref{eq77}) by $\xi^I,\xi^{II},\xi^{III},\xi^{IV}$. These vectors satisfy the following formulae:

\begin{equation}
\label{eq81}
M\xi^I=\xi^I \mu_1,\hskip5mm M\xi^{II}=\xi^{II} \mu_2,\hskip5mm M\xi^{III}=\xi^{III} \mu_3,\hskip5mm M\xi^{IV}=\xi^{IV} \mu_4.
\end{equation}

\noindent
Let us reduce the vectors $\xi^I,\xi^{II},\xi^{III},\xi^{IV}$ in the explicit form:

\noindent
\begin{equation}
\label{eq82}
\begin{array}{cc}
\xi^I=
\begin{array}{|c|}\hline
-{{y}\over{\sqrt{2r(r-z)}}}\\ \hline 
{{x}\over{\sqrt{2r(r-z)}}}\\ \hline 0\\ \hline
-{{\sqrt{r-z}}\over{\sqrt{2r}}}\\ \hline
\end{array}&
\xi^{II}=
\begin{array}{|c|}\hline
{{x}\over{\sqrt{2r(r-z)}}}\\ \hline 
{{y}\over{\sqrt{2r(r-z)}}}\\ \hline 
{{\sqrt{r-z}}\over{\sqrt{2r}}}\\ \hline
0\\ \hline
\end{array}\\&\\
\xi^{III}=
\begin{array}{|c|}\hline
{{y}\over{\sqrt{2r(r+z)}}}\\ \hline 
-{{x}\over{\sqrt{2r(r+z)}}}\\ \hline 0\\ \hline
-{{\sqrt{r+z}}\over{\sqrt{2r}}}\\ \hline
\end{array}&
\xi^{IV}=
\begin{array}{|c|}\hline
-{{x}\over{\sqrt{2r(r+z)}}}\\ \hline 
-{{y}\over{\sqrt{2r(r+z)}}}\\ \hline 
-{{\sqrt{r+z}}\over{\sqrt{2r}}}\\ \hline
0\\ \hline
\end{array}
\end{array}
\end{equation}

\noindent
Let us compile $\mu$ scalar matrix, where $\mu_1,\mu_2,\mu_3,\mu_4$ stand diagonally, along with $\xi$ matrix, where the columns are composed of 
vector components $\xi^I,\xi^{II},\xi^{III},\xi^{IV}$.

\begin{equation}
\label{eq83}
\mu=
\begin{array}{|c|c|c|c|}\hline
\mu_1&&& \\ \hline
&\mu_2&& \\ \hline
&&\mu_3& \\ \hline
&&&\mu_4\\ \hline
\end{array}
\end{equation}

\begin{equation}
\label{eq84}
\xi=
\begin{array}{|c|c|c|c|}\hline
{{-y}\over{\sqrt{2r(r-z)}}}&{{x}\over{\sqrt{2r(r-z)}}}&{{y}\over{\sqrt{2r(r+z)}}}& {{-x}\over{\sqrt{2r(r+z)}}}\\ \hline
{{x}\over{\sqrt{2r(r-z)}}}&{{y}\over{\sqrt{2r(r-z)}}}&{{-x}\over{\sqrt{2r(r+z)}}}&{{-y}\over{\sqrt{2r(r+z)}}} \\ \hline
0&{{\sqrt{r-z}}\over{\sqrt{2r}}}&0&{{\sqrt{r+z}}\over{\sqrt{2r}}} \\ \hline
{{\sqrt{r-z}}\over{\sqrt{2r}}}&0&{{\sqrt{r+z}}\over{\sqrt{2r}}}&0 \\ \hline
\end{array}
\end{equation}

\noindent
$\xi$ matrix of the form of 
(\ref{eq84}) is orthogonal. In terms of the matrices introduced (\ref{eq83}), (\ref{eq84}), the 
relations (\ref{eq81}) will be written in the form of matrix equality

\begin{equation}
\label{eq85}
M\cdot \xi=\xi\cdot \mu.
\end{equation}

\noindent
Multiply (\ref{eq85}) on the right side by $\xi^{+}$.

\begin{equation}
\label{eq86}
M=\xi\cdot\mu\cdot\xi^{+}.
\end{equation}

\noindent
If all the eigenvalues $\mu_1,\mu_2,\mu_3,\mu_4$ are positive, the relation (\ref{eq86}) can be written in the form of 
(\ref{eq50}), where 

\begin{equation}
\label{eq87}
Z=\xi\cdot\sqrt{\mu},
\end{equation}

\noindent
and the matrix 

\begin{equation}
\label{eq88}
\sqrt{\mu}=
\begin{array}{|c|c|c|c|}\hline
\sqrt{\mu_1}&&& \\ \hline
&\sqrt{\mu_2}&& \\ \hline
&&\sqrt{\mu_3}& \\ \hline
&&&\sqrt{\mu_4}\\ \hline
\end{array}
\end{equation}

\noindent
is denoted by $\sqrt{\mu}$. Let 
us write out $Z$ bispinor matrix in the explicit form for the exact solution considered.

\begin{equation}
\label{eq89}
Z=
\begin{array}{|c|c|c|c|}\hline
{{-y\sqrt{\mu_1}}\over{\sqrt{2r(r-z)}}}&{{x\sqrt{\mu_2}}\over{\sqrt{2r(r-z)}}}&{{y\sqrt{\mu_3}}\over{\sqrt{2r(r+z)}}}& {{-x\sqrt{\mu_4}}\over{\sqrt{2r(r+z)}}}\\ \hline
{{x\sqrt{\mu_1}}\over{\sqrt{2r(r-z)}}}&{{y\sqrt{\mu_2}}\over{\sqrt{2r(r-z)}}}&{{-x\sqrt{\mu_3}}\over{\sqrt{2r(r+z)}}}&{{-y\sqrt{\mu_4}}\over{\sqrt{2r(r+z)}}} \\ \hline
0&{{\sqrt{r-z}}\over{\sqrt{2r}}}\sqrt{\mu_2}&0&{{\sqrt{r+z}}\over{\sqrt{2r}}}\sqrt{\mu_4} \\ \hline
{{\sqrt{r-z}}\over{\sqrt{2r}}}\sqrt{\mu_1}&0&{{\sqrt{r+z}}\over{\sqrt{2r}}}\sqrt{\mu_3}&0 \\ \hline
\end{array}
\end{equation}

\noindent
(\ref{eq88}) for $\sqrt{\mu}$ and (\ref{eq84}) for $\xi$ orthogonal 
matrix are parent matrices to find $\Gamma_\alpha$ bispinor connectivity from (\ref{eq67}). The computing 
sequence $\Gamma_\alpha$ includes:
\begin{itemize}
\item
Formula for $\xi^{+}$, according to (\ref{eq84}) for $\xi$ matrix.
\item
Partial derivatives $\xi^{+}_{,0}, \xi^{+}_{,1}, \xi^{+}_{,2}, \xi^{+}_{,3}$.
\item
Finding $\left(\xi^{+}_{,1}\xi\right), \left(\xi^{+}_{,2}\xi\right), \left(\xi^{+}_{,3}\xi\right)$.
\item
Computation of combinations 
$\sqrt{\mu}\cdot\left(\xi^{+}_{,1}\xi\right)\cdot{{1}\over{\sqrt{\mu}}}, 
\sqrt{\mu}\cdot\left(\xi^{+}_{,2}\xi\right)\cdot{{1}\over{\sqrt{\mu}}}, 
\sqrt{\mu}\cdot\left(\xi^{+}_{,3}\xi\right)\cdot{{1}\over{\sqrt{\mu}}}$ 
and substitution in (\ref{eq67}) for $\Gamma_\alpha$.
\end{itemize}

Each of the operations enumerated is quite simple, though the treatment 
seems to be too lengthy. We give the final output introducing the auxiliary 
function

\begin{equation}
\label{eq90}
\begin{array}{c}
\Omega=\left(\sqrt{{{\mu_1}\over{\mu_3}}}+\sqrt{{{\mu_3}\over{\mu_1}}}\right)=
\left(\sqrt{{{\mu_1}\over{\mu_4}}}+\sqrt{{{\mu_4}\over{\mu_1}}}\right)=\\
=\left(\sqrt{{{\mu_2}\over{\mu_3}}}+\sqrt{{{\mu_3}\over{\mu_2}}}\right)=
\left(\sqrt{{{\mu_2}\over{\mu_4}}}+\sqrt{{{\mu_4}\over{\mu_2}}}\right)={{4mu}\over{\sqrt{4m^2 u^2 -{u'}^2}}}\\
\end{array}.
\end{equation}

\noindent
The bispinor connectivity components can be written in the compact form as 
follows:

\begin{equation}
\label{eq91}
\Gamma_0 =0,
\end{equation}

\begin{equation}
\label{eq92}
\Gamma_1 =-{{y}\over{2r(r-z)}}\cdot i\sigma_2 -\Omega\cdot{{xz}\over{4r^2 \sqrt{r^2 -z^2}}}\cdot i\rho_2 +
\Omega\cdot{{y}\over{4r\sqrt{r^2 -z^2}}}\cdot i\rho_1 \sigma_2,
\end{equation}

\begin{equation}
\label{eq93}
\Gamma_2 ={{x}\over{2r(r-z)}}\cdot i\sigma_2 -\Omega\cdot{{yz}\over{4r^2 \sqrt{r^2 -z^2}}}\cdot i\rho_2 -
\Omega\cdot{{x}\over{4r\sqrt{r^2 -z^2}}}\cdot i\rho_1 \sigma_2,
\end{equation}

\begin{equation}
\label{eq94}
\Gamma_3 =\Omega\cdot{{\sqrt{r^2 -z^2}}\over{4r^2}}\cdot i\rho_2 .
\end{equation}

\noindent
As we might expect, the formulae (\ref{eq91})-(\ref{eq94}) obtained correspond to the group 
of gauge generations $SO(4)$.

\subsection{Solution analysis}
\label{subsec:solution}

Let us write the formulae for energy-momentum tensor components 
corresponding to the solution of (\ref{eq75}). Suppose that the values included in 
energy-momentum tensor are of the structure as follows:

\begin{equation}
\label{eq95}
A_\alpha =\left(A_0 (x,y,z),0,0,0\right),\hskip10mm \lambda=Const.
\end{equation}

\noindent
We have:

\begin{equation}
\label{eq96}
\left.
\begin{array}{c}
T_{00}=-3(A_0 )^2 -\lambda\\
T_{0k}=A_{0,k}\\
T_{mn}=\delta_{mn}\left[(A_0 )^2 +\lambda\right]
\end{array}
\right\}
\end{equation}
\noindent
Change $A_\alpha,\lambda$ in (\ref{eq96}) to 
$J_\alpha,m$ according to the formulae (\ref{eq35}), (\ref{eq36}). We get:

\begin{equation}
\label{eq97}
\left.
\begin{array}{c}
T_{00}=-3m^2 (J_0 )^2 -2m^2\\
T_{0k}=m J_{0,k}\\
T_{mn}=\delta_{mn}\left[m^2 (J_0 )^2 +2m^2\right]
\end{array}
\right\}
\end{equation}

\noindent
(\ref{eq97}) implies the formulae to express $U$ energy density and $P$ pressure

\begin{equation}
\label{eq98}
U=-3m^2 (J_0 )^2 -2m^2,
\end{equation}

\begin{equation}
\label{eq99}
P=m^2 (J_0 )^2 +2m^2.
\end{equation}

It follows from (\ref{eq98}), (\ref{eq99}) that the energy density is the negative value, 
and the pressure is positive,

\begin{equation}
\label{eq100}
U<0,\hskip10mm P>0.
\end{equation}
\noindent
Here $U$ and $P$ are related as

\begin{equation}
\label{eq101}
U=-3P+4m^2 .
\end{equation}

\noindent
If we compare (\ref{eq101}) and (\ref{eq99}) we get the following relation

\begin{equation}
\label{eq102}
P={{1}\over{16V^2 m^4}}+2m^2.
\end{equation}
\noindent
The substitution of (\ref{eq102}) for (\ref{eq101}), gives:

\begin{equation}
\label{eq103}
U=-{{3}\over{16V^2 m^4}}-2m^2.
\end{equation}

\noindent
The comparison of (\ref{eq103}) and (\ref{eq98}) gives:

\begin{equation}
\label{eq104}
J^0
={{1}\over{4V m^3}}.
\end{equation}

\noindent
It follows from the general theory that the vectors $J^\alpha$ and $u^\alpha$ should be collinear, i.e.

\begin{equation}
\label{eq105}
J^\alpha =Const\cdot {{u^\alpha}\over{V}}.
\end{equation}

\noindent
By taking the index $\alpha=0$ in (\ref{eq105}) and using the formulae (\ref{eq104}) and (\ref{eq75}) for 
$J^0$, we get:

\begin{equation}
\label{eq106}
\rho={{1}\over{V}}=4cm^3 \cdot {{e^{2mr}}\over{mr}}.
\end{equation}

\noindent
(\ref{eq106}), (\ref{eq103}), (\ref{eq102}) give us radial coordinate dependence of energy density 
and pressure:

\begin{equation}
\label{eq107}
U=-3c^2\cdot{{e^{4mr}}\over{r^2}}-2m^2,
\end{equation}

\begin{equation}
\label{eq108}
P=c^2\cdot{{e^{4mr}}\over{r^2}}+2m^2.
\end{equation}
\noindent
It follows from (\ref{eq107}), (\ref{eq108}) that when $r\rightarrow \infty$, $U$ and $P$ are 
related as $U= -P$. 

Now let us find the formula for $c_s$, isentropic velocity of sound. Substitute (\ref{eq108}) 
and (\ref{eq106}) for $P$ and $1/V$ in (\ref{eq24}). We get:

\begin{equation}
\label{eq109}
c^2_s ={{1}\over{6}}\cdot c\cdot{{e^{2mr}}\over{m^2 r}}.
\end{equation}

\noindent
(\ref{eq109}) states that isentropic velocity of sound goes to infinity when 
$r\rightarrow\infty$. Such dependence 
implies that the velocity of sound should come up to the velocity of light 
at certain finishing radius $\overline{r}$. In order for $\overline{r}$ to be found, (\ref{eq109}) should be set equal to 
$1/m$ - the unique constant in length dimension problem.

\begin{equation}
\label{eq110}
\left(m\overline{r}\right)e^{-2m\overline{r}}={{6}\over{11}}\cdot c.
\end{equation}

\noindent
It is clear that if

\begin{equation}
\label{eq111}
c\leq 11/12e,
\end{equation}

\noindent
the solution of (\ref{eq110}) is always available. The perturbation velocity of 
geometrodynamic medium does not achieve the velocity of light. 

Evidently, the accumulating perturbations reconstruct the solution within 
the framework of CGD solutions if the character of solution formation is 
evolutionary in the range of radii near $\overline{r}$. In other words, $\overline{r}$ - 
is the radius within which the solution can 
branch, i.e. one solution branch can be changed by the other. Note that the 
solution of (\ref{eq74}) type can be the other solution branch with decreasing 
exponent not increasing.

Thus, the solution (\ref{eq70}) is valid for all values of the variable 
$r>0$. However, there are 
two values of the variable when the solution and/or its treatment undergo 
changes. The first value is $1/4m$, and the second value $\overline{r}$ is determined from (\ref{eq110}). The full solution includes 
the description of $J_\alpha$ vector and $H_{\alpha\beta}$ tensor in three domains $I,II,III$. Domain $II$ is

\begin{equation}
\label{eq112}
(1/4m)<r<{\overline{r}} ,
\end{equation}
\noindent
it differs in the fact that if the integration constant satisfies the inequality (\ref{eq111}), 
the condition of positivity of the polarization matrix of $M$ density from (\ref{eq53}) is fulfilled in the 
range (\ref{eq112}). It means that the solution (\ref{eq70}) can be interpreted in terms of 
$Z$ bispinor matrix (\ref{eq89}) 
in the range of (\ref{eq112}). Because this matrix is the direct sum of four 
bispinors, the solution can be described in terms of four particles with 
$1/2$ spin in the domain $II$. As for the domains 
$I, III$, the polarization matrix of $M$ density from (\ref{eq53}) is positively definite in these domains. 
$M$ matrix requires additional analysis in these domains.

\section{Result discussion }
\label{sec:mylabel3}

The work is one of the fundamental areas of the modern theoretical physics 
-- determination of physical meaning of Weyl space-time degrees of freedom. 
The problem emerged in 1918 when H.Weyl [7] 
suggested considering general space in relativity theory not Riemannian 
space. The additional space properties were connected with the vector 
introduced (Weyl vector). Physicists and mathematicians have been clarifying 
the meaning of the vector for more than 90 years. Partial success attends 
the research confirming the assumption of the fundamental function of Weyl 
vector in physics. Thus, the conceptual models appeared in which Weyl 
degrees of freedom are connected with parameters of dark material and energy 
in the Universe, with cosmological red shift, with change of scale to 
measure the space-time interval ([8], [14]-[19], etc.). In some works ([20]-[22], etc.) 
Weyl degrees of freedom are considered as Weyl integrable space quality 
(i.e. the space in which Weyl vector is the gradient of scalar function) 
resulting in Schrodinger equation. 

The present work suggests the solution to the problem of physical 
interpretation of Weyl space degrees of freedom. In terms of CGD, Weyl 
vector depends on the range of the phenomena considered. Weyl vector acts as 
current density vector of conserved charge on a large scale, i.e. in case 
the left side of the equation (\ref{eq1}) with the right side in the form of tensor 
(\ref{eq2}) cannot be omitted and the space curvature cannot be neglected. CGD does 
not predefine the type of conserved charge, but shows that this charge does 
exist. The availability of such a strongly conserved current density vector 
allows introduction of specific volume notion, as well as get up 
phenomenological thermodynamics for geometrodynamic continuum. 

Weyl vector is proportional to sum vector of probability density of all the 
particles with half-integer spin on microparticle scale in terms of which 
the geometrodynamic medium dynamics can be described. The gauge is always 
available when the divergence of Weyl vector goes to zero.

It is our opinion that the agreement of CGD equations with Dirac derived 
equations given in Table \ref{tab2}, is the trenchant 
argument confirming the viability of interpretation suggested. 

However, it should be understood that the dynamics of Weyl degrees of 
freedom is described by (\ref{eq31})-(\ref{eq34}) at any polarization structure of 
$M$ matrix. However, 
quantum-field interpretation of these equations is not always possible, it 
is applicable when all the eigenvalues of $M$ matrix are positive. This condition fits the 
requirements imposed on polarization density matrices in quantum mechanics 
and quantum field theory. Evidently, the violation of positivity condition 
of $M$ matrix, as well 
as vanishing of matrix determinant do not interfere with the analogous 
interpretation -- it is a matter that requires additional analysis. The 
results of the work assume the number of generalizations, for example: 
complexification, introduction of internal spaces, various cases of 
$M$ matrix eigenvalues, etc. 

We note finally that Weyl vector interpretation described opens up new 
possibilities when Standard model of elementary particles, particularly 
confinement model, is theoretically proved.

\label{sec:referencesnally}

\end{document}